\documentclass[namedreferences,hyperref,optionalrh,solaromanenum]{spr-sola}

\usepackage{graphicx,times}                    
\usepackage{amssymb, amsmath}                    
\usepackage{color}                       

\graphicspath{{figures/}}

\chardef\us=`\_

\begin{document}

\begin{frontmatter}

\title{The role of photospheric magnetic flux diffusion in initiation of solar eruptions}


\author[addressref={aff1,aff3},email={bianxk@hit.edu.cn}]{\inits{X.}\fnm{Xinkai}~\snm{Bian}\orcid{0000-0001-9189-1846}}
\author[addressref={aff1},corref,email={chaowei@hit.edu.cn}]{\inits{C.}\fnm{Chaowei}~\snm{Jiang}\orcid{0000-0002-7018-6862}}
\author[addressref={aff3},corref,email={ywangsz@hit.edu.cn}]{\inits{Y.}\fnm{Yang}~\snm{Wang}\orcid{0000-0002-4581-9242}}
\author[addressref={aff1},corref]{\inits{Zou1}\fnm{Peng}~\snm{Zou}\orcid{0000-0002-8474-0553}}
\author[addressref={aff1},corref]{\inits{Feng}\fnm{Xueshang}~\snm{Feng}\orcid{0000-0001-8605-2159}}
\author[addressref={aff1},corref]{\inits{Zuo1}\fnm{Pingbing}~\snm{Zuo}\orcid{0000-0003-4711-0306}}
\author[addressref={aff1},corref]{\inits{Wang1}\fnm{Yi}~\snm{Wang}\orcid{0000-0002-7094-9794}}

\address[id=aff1]{State Key Laboratory of Solar Activity and Space Weather, School of Aerospace, Harbin Institute of Technology, Shenzhen 518055, China}
\address[id=aff3]{School of Science, Harbin Institute of Technology, Shenzhen 518055, China}
%
\runningauthor{Bian et al.}
\runningtitle{Initiation mechanism of solar eruptions}

\begin{abstract}
	Solar eruptions may occur at different evolutionary stages of active regions, during which the photospheric motions manifest in various forms, including flux emergence, sunspot rotation, shearing, converging, and magnetic flux diffusion. However, it remains unclear what are the specific roles played by these different motions in leading to eruptions. 
	Here, we employ high resolution magnetohydrodynamic simulations to demonstrate how solar eruptions can be initiated in a single bipolar configuration, driven by first shearing and then flux diffusion at the bottom surface.
	Flux diffusion disperses the photospheric magnetic flux, driving portions of it toward the polarity inversion line (PIL). This process leads to the expansion of core field, enhancing the pinching effect to form the current sheet. When magnetic reconnection occurs within this current sheet, the eruption is initiated, characterized by a rapid release of magnetic energy and accompanied by the formation of a erupting flux rope.
	Additionally, flux diffusion contributes to magnetic cancellation near the PIL, leading to the formation of a weakly twisted magnetic flux rope prior to the eruption. However, this pre-exist flux rope plays a limited role in eruption initiation, as its spatial position remains largely unchanged throughout the eruption.
	These findings demonstrate that the primary role of flux diffusion is to facilitate current sheet formation, highlighting the critical role of current sheet formation in eruption initiation.
\end{abstract}

\end{frontmatter}

\section{Introduction}\label{sec:intro}

Solar eruptions, including flares and coronal mass ejections (CME), are among the most violent form of energy release in the heliosphere. 
These events are primarily powered by the magnetic energy stored in the solar corona~\citep{Priest2002, Forbes2006, Fleishman2020}.
Due to the extremely high temperatures in the corona, reaching several million Kelvin, resulting in a plasma $\beta$ much less than unity. In this low-$\beta$ environment, the magnetic field dominates the plasma dynamics.
In contrast, the photosphere has a much higher plasma density and a correspondingly high plasma $\beta$, meaning that plasma motions dominates the magnetic field~\citep{Gary2001}.

The footpoints of coronal magnetic field lines are anchored in the photosphere. As a result, various types of photospheric surface motion, such as shearing along the polarity inversion line (PIL), sunspot rotation, flux converging toward the PIL, and flux dispersion, continuously drive the evolution of the coronal magnetic system, leading to the gradual accumulation of magnetic energy~\citep{Brown2003, VanDriel-Gesztelyi2003, Min2009, Vemareddy2012, Lamb2013, Vemareddy2017, Chintzoglou2019}. This process is commonly referred to as the line-tied effect~\citep{Aulanier2016}.

During this energy accumulation phase, the coronal magnetic field typically evolves in a quasi-static, force-free equilibrium.
However, there exists an upper limit to the amount of magnetic energy that can be stably stored in the corona. Once the magnetic energy approaches a critical threshold~\citep{ Moore2012, Sun2015}, the system becomes unstable and must release the stored energy through an eruption. The physical process that breaks the coronal equilibrium and initiates magnetic energy release is referred to as the triggering mechanism of solar eruptions.
Currently, numerous theoretical models have been proposed based on observations~\citep{Forbes2000, Gopalswamy2004, Hudson2006, Chen2011, Hudson2011, Webb2012, Schmieder2012, Janvier2015, Cheng2017, Green2018, Zhang2021, Jiang2024a}. However, the exact trigger for each individual event remains case-dependent and requires detailed analysis.

Active regions (ARs) serve as the primary sources of solar eruptions, which are the magnetized areas extending from the photosphere to the corona. 
While eruptions can occur at any stage of AR evolution, they are most frequent during the emergence stage and decreasing thereafter~\citep{VanDriel-Gesztelyi2015}. 
During the decaying phase of ARs, the flux diffusion (or dispersion) becomes the dominant form of motion. However, how solar eruptions are initiated in decaying ARs remains an open question.

Previous simulation studies has suggested that flux diffusion and magnetic cancellation solely contribute to the formation of twisted magnetic flux rope (MFR)~\citep{VanBallegooijen1989, Amari2003, Amari2007, Amari2010, Aulanier2010}.
Once formed, the MFR gradually rises from a position near the photosphere under the bottom driving. It remains stable throughout this process. However, when MFR reaches a critical height, the background magnetic field can no longer provide sufficient confinement to maintain its equilibrium, that is, the torus instability causes the eruption~\citep{Kliem2006}. 
Since both flux diffusion and converging motions pull the footpoints of partial magnetic field lines toward the PIL, some simulation studies adopted magnetic field convergence and magnetic cancellation to drive the system's evolution and found a similar process of MFR formation, which ultimately led to an eruption triggered by ideal MHD instability~\citep{Amari2003a, Amari2011a, Zuccarello2015, Hassanin2022}.

However, in our previous simulation, we apply converging motion without magnetic cancellation. This approach increases the magnetic gradient across the PIL without magnetic cancellation. The converging motion at the footpoints of the sheared arcade creates a current sheet in a quasi-static way, and the eruption is subsequently triggered by magnetic reconnection within that current sheet~\citep{Bian2022}.
Based on this, we propose that flux diffusion not only facilitates the MFR formation, but also contributes to the development of current sheet within sheared magnetic field.
Likewise, we propose that magnetic reconnection within the current sheet serve as the primary mechanism for both triggering and driving the eruption~\citep{Jiang2021b, Bian2022a, Jiang2024}, rather than MHD ideal instabilities of a pre-existing flux rope.

In this study, we employ high-resolution three-dimensional magnetohydrodynamic (MHD) numerical simulations to investigate the role of magnetic flux diffusion in solar eruption initiation. For a sheared coronal magnetic field configuration, flux diffusion at the bottom boundary effectively mimics converging motions, drawing the footpoints of magnetic field lines toward the PIL. This process results in magnetic cancellation at the PIL and the formation of weakly twisted magnetic flux ropes. However, our results indicate that this flux rope does not contribute to the triggering process, as it remains below the flare current sheet and undergoes minimal height change before and after the eruption.
Crucially, our simulations demonstrate that flux diffusion facilitates the formation of a current sheet, and that magnetic reconnection within this current sheet is the key process responsible for both triggering and driving the eruption.

This paper is organized as follows: Section~\ref{sec:methods} present the MHD model, including the initial and boundary conditions, with a focus on surface diffusion. Then, the simulation results are described in Section~\ref{sec:res}. Finally, Section~\ref{sec:discussion} provides a discussion and summarizes our conclusions.

\section{Method}
\label{sec:methods}
Our model solves the 3D MHD equations in a Cartesian coordinate system using the conversation element and solution element (CESE) scheme~\citep{Feng2010, Jiang2010, Jiang2016a, Jiang2021b}. Parallel computations are performed using the FORTRAN-based Paramesh package~\citep{MacNeice2000}. The basic configuration of the model remains consistent with our previous study~\citep{Jiang2021b}.
Notably, to minimize artificial resistivity, we set explicit resistivity to zero in the magnetic induction equation throughout the simulation domain, except in regions influenced by bottom-surface diffusion. As a result, magnetic reconnection occurs solely through numerical diffusion when the thickness of the current layer approaches the grid resolution. This ensures that reconnection processes are physically governed by the numerical scheme rather than artificially imposed resistivity.

We employ the dimensionless parameters for normalization.
The normalized values of magnetic field, length and time are $1.86$~G, $11.52$~Mm and $105$~s, respectively, resulting in a corresponding normalized velocity of $110$~km~s$^{-1}$. Hereafter, all physical quantities are expressed in normalized values, ensuring consistency and facilitating direct comparison of results.
The numerical simulations are conducted within a cubic computational domain of $48^3$. The computational grid is block-structured using adaptive mesh refinement (AMR) scheme, with a base-level resolution of $\Delta x=\Delta y=\Delta z=0.25$. To accurately capture key physical processes, such as magnetic cancellation and current sheet formation, the maximum refinement level achieves a resolution of $\Delta =7.8125\times 10^{-3}$.

The initial magnetic field is constructed using a photospheric magnetogram described by the analytical expression:
\begin{equation}\label{eq:magnetogram}
	\begin{split}
		B_{z}(x,y,0) = B_0
		e^{-x^2/\sigma_x^2}(e^{-(y-y_c)^2/\sigma_y^2}-e^{-(y+y_c)^2/\sigma_y^2}).
	\end{split}
\end{equation}
Here, $B_0=20$ represents the characteristic magnetic field strength. The spacial distribution parameters $\sigma_x=2.5$ and $\sigma_y=1.25$ define the extent of magnetic polarity in the $x-$ and $y-$ directions, respectively, while $y_c=1$ represents the separation distance between the centers of the opposite magnetic polarities.

To investigate the role of magnetic flux diffusion in eruption initiation, we first apply photospheric velocity driving at the bottom, allowing the magnetic field to develop magnetic shear (and an amount of magnetic free energy) without triggering an eruption. Subsequently, we halt the velocity driving and introduce magnetic flux diffusion to further guide the system's evolution.
In our simulation, the photospheric driving motion mimics the rotational motion of sunspots and is implemented through a velocity field derived from the magnetic flux distribution at the bottom boundary. This velocity effectively injects free magnetic energy into the system while inducing magnetic shear near the PIL, without modifying the magnetic flux distribution. The velocity components are defined as follows,
\begin{equation}\label{eq:dirven_speed}
	\begin{split}
		v_{x}=\dfrac{\partial \psi(B_{z})}{\partial y};
		v_{y}=-\dfrac{\partial \psi(B_{z})}{\partial x},
	\end{split}
\end{equation}
where the function $\psi$ is given by,
\begin{equation}\label{eq:dirven_speed_psi}
	\begin{split}
		\psi = v_{0}B_{z}^{2}e^{-(B_{z}^{2}-B_{z, {\rm max}}^{2})/B_{z, {\rm max}}^{2}}.
	\end{split}
\end{equation}
The maximum driven velocity is set to $0.04$, while remains significantly below the local Alfv$\acute{\text{e}}$nic ($>10$), ensuring that the driving process maintains quasi-static evolution. 
All initial and boundary conditions in our simulations are consistent with those described in~\citet{Jiang2021b}. For further details, please refer to that paper.

Since flux diffusion cannot be adequately represented by simple photospheric velocity motions, we introduce a photospheric diffusion coefficient into the magnetic induction equation to simulate the diffusion process, following many other studies (e.g., \citealp{Aulanier2010, Rice2023}),
\begin{equation*}\label{eq:mag_induc}
	\dfrac{\partial \textbf{B}}{\partial t}= \nabla\times(\textbf{v} \times \textbf{B}) + \eta_{ \text {pho}} \nabla_{\perp}^2 \textbf{B}.
\end{equation*}
Here, the diffusion coefficient $\eta_{ \text{pho}}$ is set to $2 \times 10^{-3}$, corresponding to a characteristic timescale of $500$. This is much longer than the Alfv$\acute{\text{e}}$nic timescale of $0.1$ in the bottom magnetic field region, associated with the local Alfv$\acute{\text{e}}$nic velocity of $10$. This small coefficient ensures that the diffusion process remains quasi-static throughout the magnetic field evolution.

In this study, we conduct two sets of simulations, both involving bottom velocity driving until $t=100$. Each simulation starts with an initial potential field, and the magnetic field evolution during the energy injection phase (i.e., bottom surface driving) follows the same process as described in previous studies~\citep{Jiang2021b, Bian2022a, Bian2022}.
Beyond this point, the simulations evolve in different ways, one continues under sustained velocity driving, while the other transitions into a relaxation phase at $t=100$, followed by diffusion driving starting at $t=150$. 
These two simulations are labeled as cases $E$ and $D$, respectively.

\section{Results}
\label{sec:res}

\begin{figure}
	\centering
	\includegraphics[width=1.0\textwidth]{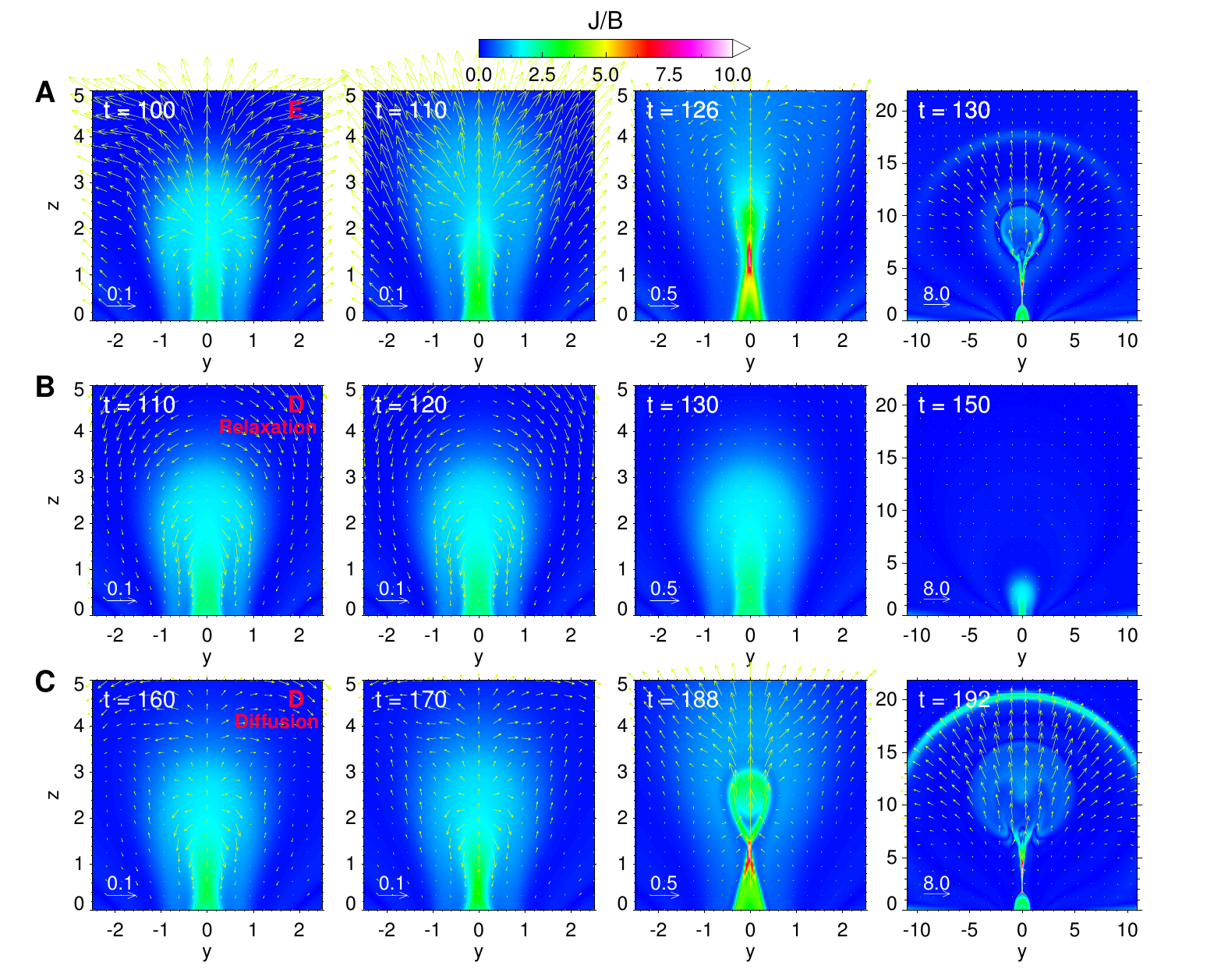}
	\caption{Comparison of the current distribution and surface velocity on the central vertical slice (i.e., the $x=0$ plane) for two simulation cases in different driving phase. (A), (B) and (C) represent the simulation cases $E$, $D$ in relaxation phase and $D$ in diffusion phase, respectively. } 
	\label{fig:compare}
\end{figure}

\begin{figure}
	\centering
	\includegraphics[width=0.8\textwidth]{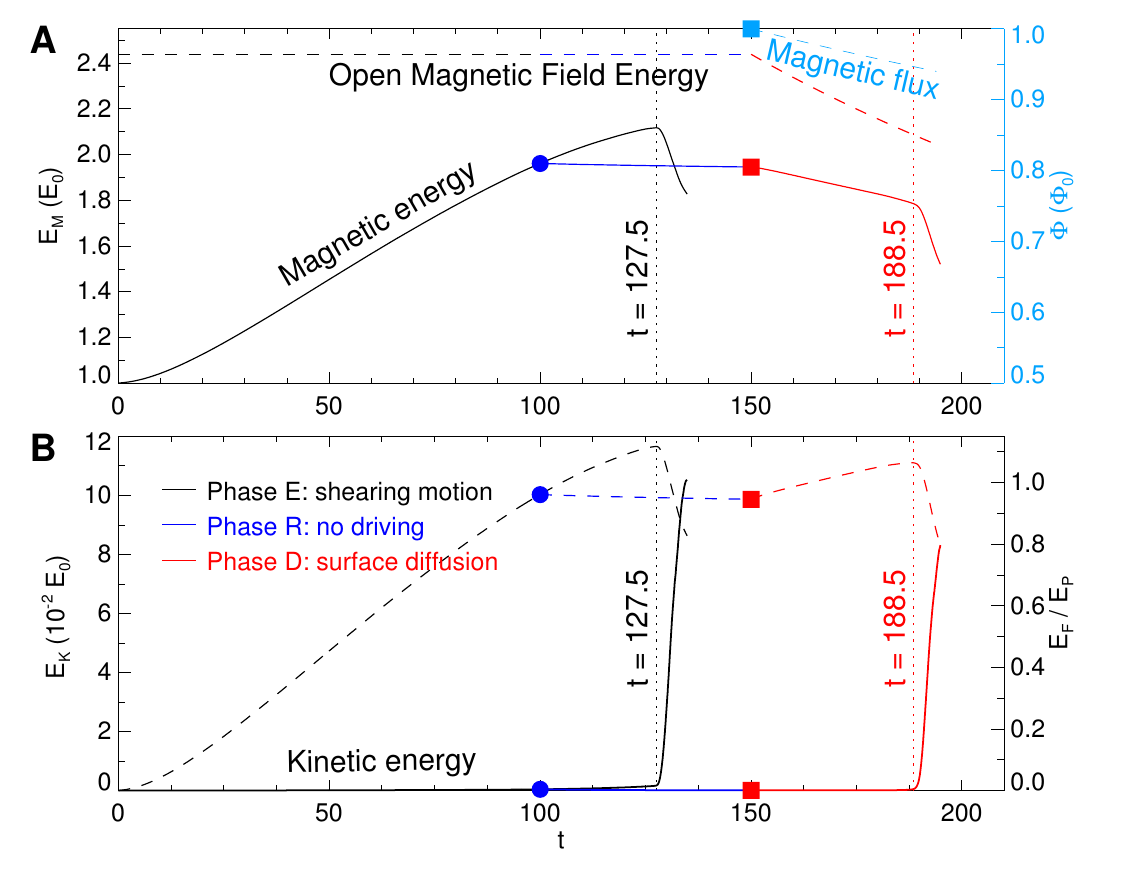}
	\caption{Temporal evolution of magnetic and kinetic energy. (A) Magnetic energy evolution in different driven stage. The black, blue and red colors indicate the energizing, relaxation and surface diffusion, respectively. The circle and square indicate the begin of relaxation and surface diffusion, respectively. The dashed line indicates the open field energy corresponding to the magnetic flux distribution on the bottom surface. The vertical dotted lines show the onsets of the eruption. The sky dashed line indicates the normalized magnetic flux, i.e. $\Phi/\Phi_0$ in the bottom surface during the diffusion phase. (B) Kinetic energy (solid line) and non-potentiality  $E_F/E_P$ (dashed line) evolution. } 
	\label{fig:energy}
\end{figure}

In simulation $E$, where bottom driving continuous beyond $t=100$, the core magnetic structure gradually evolves into a strongly sheared, inverse S-shape configuration. Meanwhile, the current within the core field is progressively compressed and strengthened, ultimately leading to the formation of a current sheet. 
When the thickness of this current sheet decreases to the grid resolution (at $t=123$), numerical resistivity comes into play, triggering magnetic reconnection (Figure~\ref{fig:compare}A). 
It should be noted that magnetic reconnection within the current sheet is resolution dependent. In higher resolution simulations, the eruption onset is delayed because the current sheet takes longer to thin to the grid resolution where the numerical resistivity becomes effective.
Consequently, a magnetic flux rope forms via reconnection, rapidly rising and expanding. A detailed analysis of the acceleration of magnetic field lines confirms that magnetic reconnection plays a dominant role in both triggering and driving the eruption~\citep{Jiang2021b}. 
The energy evolution in simulation $E$ can be divided into two distinct phases: a quasi-static energy injection phase, followed by an eruption phase characterized by a rapid decrease in magnetic energy, as illustrated by the black curve in Figure~\ref{fig:energy}A.

\begin{figure} 
	\includegraphics[width=1.0\textwidth]{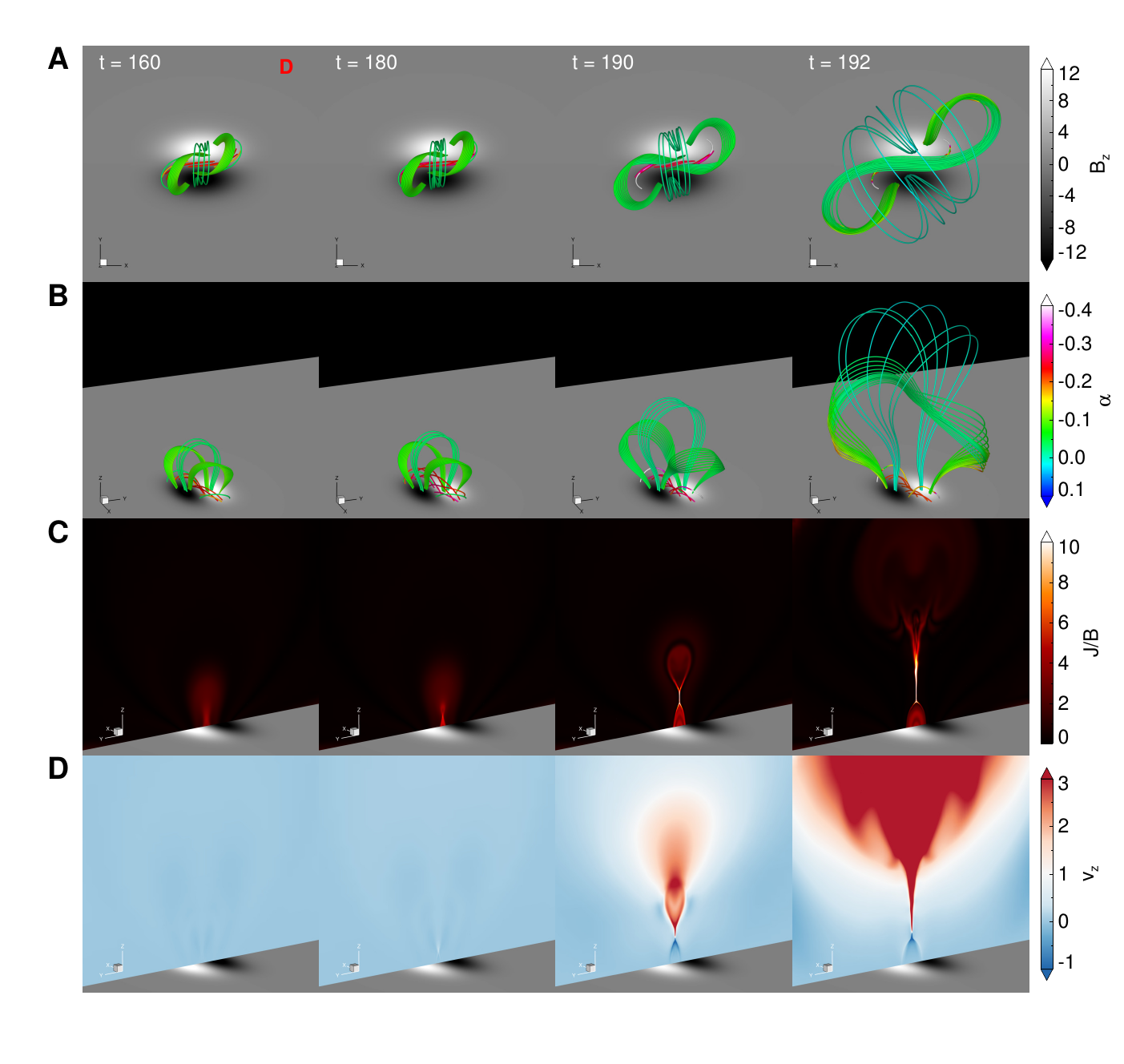}
	\centering
	\caption{Evolution of magnetic field lines, current and velocity in the simulation case $D$. (A) 2D view of the magnetic field lines. The colored lines represent the magnetic field lines, and the colors denote the value of the nonlinear force-free factor $\alpha$. The background shows the magnetic flux distribution on the bottom boundary. (B) 3D perspective view of the same field lines shown in panel A. (C) Distribution of normalized current $J/B$ on the central vertical slice. (D) Velocity on the same slice as shown in panel C.} 
	\label{fig:mag}
\end{figure}

In simulation $D$, velocity driving is halted at $t=100$, prior to the eruption onset, allowing the system to evolve freely. The overall magnetic field structure and current density remain largely unchanged~(\ref{fig:compare}B), and the magnetic energy (blue curve in Figure~\ref{fig:energy}) remains nearly constant.
Then, bottom diffusion is activated at $t=150$, leading to the gradual outward diffusion of photospheric magnetic flux~(cyan curve in Figure~\ref{fig:energy}A). 
In this process, the core flux can expand for pinching to form the current sheet, the evolution of the current density are similar to simulation $E$, as shown in Figure~\ref{fig:compare}A and C.   
Figure~\ref{fig:mag} provides a comprehensively illustration of the temporal evolution of key system's features during the diffusion driving phase, including magnetic field lines, electric current distribution, and velocity field. 

\begin{figure} 
	\includegraphics[width=0.8\textwidth]{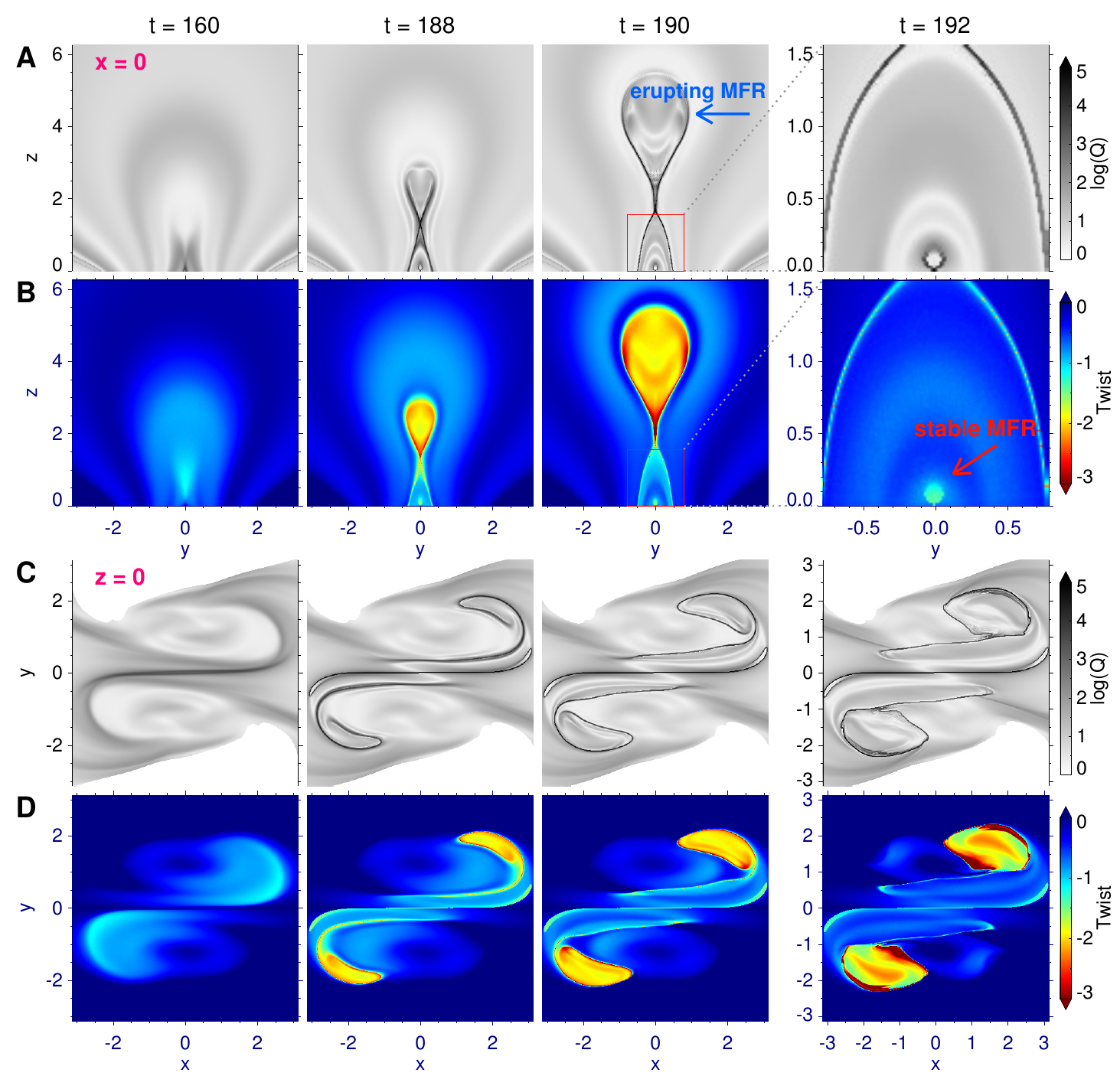}
	\centering
	\caption{Magnetic topology evolution on the central vertical slice and bottom surface. (A) and (C) show magnetic squashing degree $Q$. (B) and (D) show magnetic twist number $T_{w}$. The blue and red arrows represent the newly formed MFR during the eruption and the pre-existing MFR, respectively.} 
	\label{fig:qsl}
\end{figure}

Figure~\ref{fig:qsl}A shows the spatial distribution of the magnetic squashing degree $Q$ along the central slice. This metric quantifies the gradient of magnetic field line mapping with respect to their footpoints and is essential for identifying quasi-separatrix layers (QSLs) of magnetic fields~\citep{Titov2002}.
At $t=188$, a distinct X-shaped QSL appears, marking the region most susceptible to magnetic reconnection, i.e., the current sheet.
The eruption is initiated when magnetic reconnection occurs within this current sheet at $t=188.5$.

During the eruption phase, a magnetic flux rope (erupting MFR) forms through reconnection and rapidly rises and expands~(Figure~\ref{fig:mag}B and \ref{fig:qsl}). 
Simultaneously, magnetic energy decreases sharply, while kinetic energy increases significantly, reaching nearly 10$\%$ of the initial potential energy, as shown in Figure~\ref{fig:energy} red curve. 
The reconnection within the current sheet triggers the eruption, i.e., the rapid release of magnetic energy, consistent with the triggering mechanism identified in Simulation $E$.

The dotted line in Figure~\ref{fig:energy}A shows the evolution of open force-free field energy, which is defined by the following equation mentioned in \cite{Amari2003a} and calculated based on the vertical component of the bottom magnetic field,
\begin{equation}\label{eq:open_energy}
	\begin{split}
	E_{\rm{open}}(t) = \frac{1}{{16{\pi ^2}}} \int_{{S_1} \times {S_2}} {\frac{{\left| {{B_z}({x_1},{y_1},0,t){B_z}({x_2},{y_2},0,t)} \right|}}{{\sqrt {{{({x_2} - {x_1})}^2} + {{({y_2} - {y_1})}^2}} }}{\rm{d}}{S_1}{\rm{d}}{S_2}}.
	\end{split}
\end{equation}
This represents the upper energy limit for all force-free field and simply connected fields with a fixed flux distribution, as conjectured by Aly and Sturrock~\citep{Aly1991, Sturrock1991}. 
As the magnetic flux decreases, the potential energy, total magnetic energy, and open field energy all decrease accordingly. 
Although the potential energy and open field energy decline at a rate comparable to that of the magnetic flux, the total magnetic energy decreases more slowly than both. 
As a result, during the diffusion phase, the ratio of magnetic energy to open field energy gradually increases, reflecting an enhanced degree of field non-potentiality. At the eruption onset ($t=188.5$), this ratio is approximately $0.85$, which is consistent with the value obtained in simulation $E$ at its eruption onset ($t=127.5$). We also present the magnetic non-potentiality, $E_F/E_P$, in Figure~\ref{fig:energy}B. At the eruption onset of both simulations, this ratio reaches about $1.1$, indicating a comparable degree of departure of the magnetic field from its potential state. This clearly demonstrates the existence of a magnetic energy threshold for a given flux distribution, with eruption occurring once this threshold is approached~\citep{Bian2022b}. 
Similarly, higher resolution simulation tend to yield a slightly higher ratio, as more free magnetic energy can accumulate prior to eruption, however, the increase remains modest.

\begin{figure} 
	\includegraphics[width=0.8\textwidth]{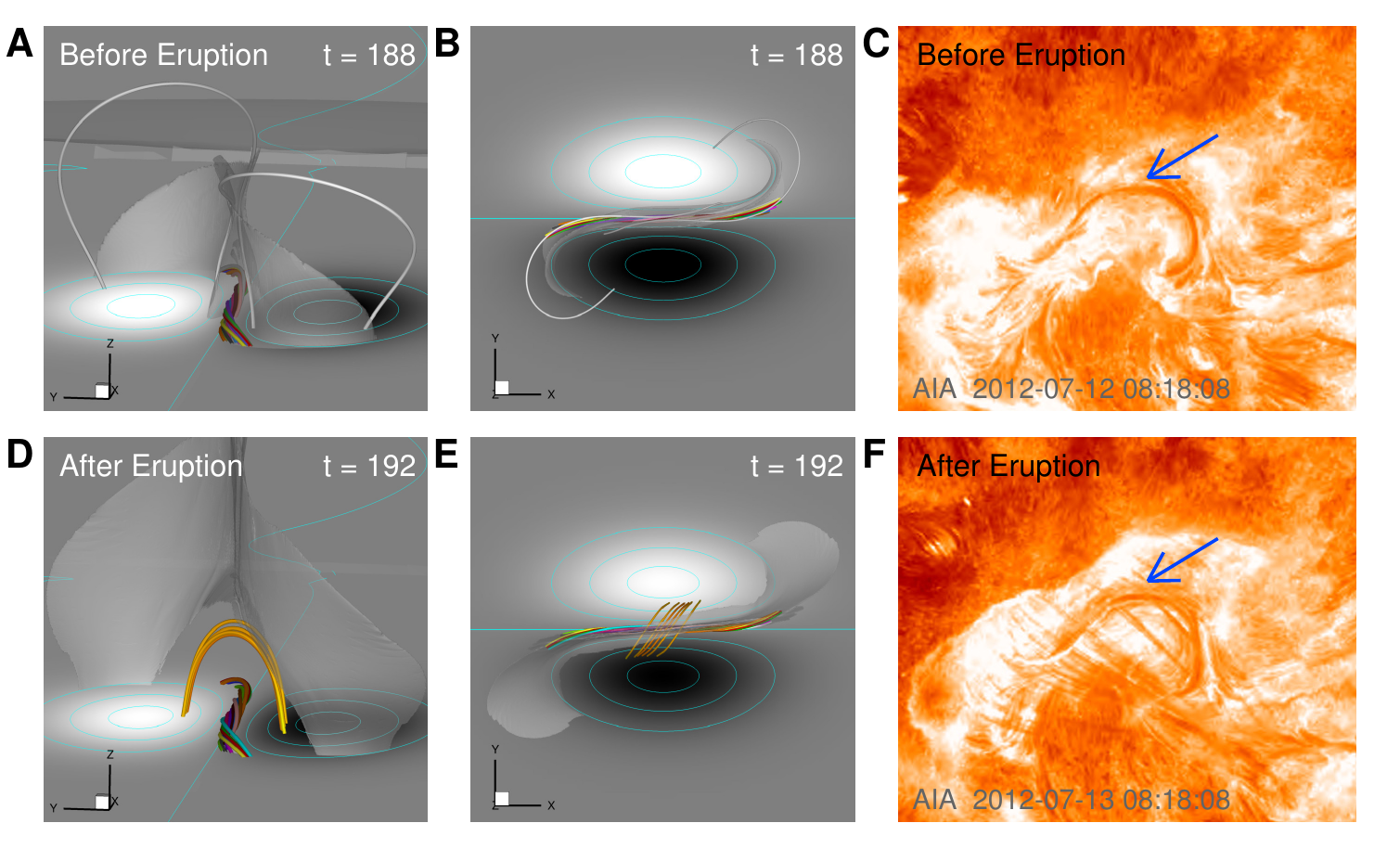}
	\centering
	\caption{Magnetic configuration in the simulation $D$ and observational images before and after the flare. (A), (B), (D) and (E) show the simulation results, where the colored lines represent the pre-existing MFR. The current sheet in a 3D configuration is illustrated by the iso-surface of $J/B=5.75$ and $J/B=85$, respectively, while the white lines indicate the magnetic field lines where magnetic reconnection is expected to occur. The orange lines in panels (D) and (E) represent the post-flare loops. (C) and (F) $SDO/$AIA $304$ \AA $ $ observed filament structure during the flare.} 
	\label{fig:MFR}
\end{figure}

A distinct feature of simulation $D$ is the bottom distribution of the magnetic squashing factor $Q$, which provides clear evidence that magnetic diffusion drives magnetic cancellation along the PIL. 
Analysis of the magnetic twist number $T_{w}$~\citep{Berger2006} reveals that this cancellation process facilitates the formation of another twisted flux rope (stable MFR), distinct from the erupting one. 
This MFR appears at $t=188$ before the eruption onset, and it characterized by its low altitude ($z<0.5$) and moderate magnetic twist of approximately one turn, as clearly shown in Figure~\ref{fig:qsl}B and D. This structure aligns well with the theoretical descriptions of MFR formation through the magnetic cancellation mechanism~\citep{VanBallegooijen1989}.
Notably, this MFR remains stable near the bottom boundary through the eruption~(Figure~\ref{fig:MFR}A and D). Similar phenomena have been identified in various observed events, such as the X$1.4$-class flare in AR 11520, which began at 16:10 UT on 2012 July 12~\citep{Cheng2014}. 
In this event, the filament structure remained stable in both position and configuration, while the flare loops formed directly above it (Figure~\ref{fig:MFR}C and F). This observational behavior resembles the characteristics reproduced in our simulation (Figure~\ref{fig:MFR}B and E), implying that the flux rope play a limited role in the eruption initiation process.

\section{Discussions and Conclusion}
\label{sec:discussion}

Previous numerical studies have suggested that magnetic flux diffusion and magnetic cancellation play a crucial role in the formation of twisted flux rope. Within this framework, flux ropes are generally assumed to pre-exist before eruption onset, with their triggering primarily governed by ideal MHD instabilities~\citep{Amari2003, Amari2003a, Kliem2006, Amari2010, Aulanier2010, Amari2011a, Zuccarello2015, Hassanin2022}.
However, we propose that the diffusion process may function similarly to converging motions, which draw the footpoints of magnetic field lines toward the PIL, effectively compressing the current layer within the sheared magnetic field. This compression facilitates the formation of a current sheet, ultimately triggering eruptions through magnetic reconnection~\citep{Bian2022}. 	
Notably, our high-resolution MHD simulations exhibit several distinct features compared to previous studies. First, they transfer a substantially smaller amount of magnetic flux into the flux rope, which may account for its relatively lower height. Second, the MHD equations are solved using the CESE scheme, which is inherently characterized by low numerical dissipation. Moreover, no explicit magnetic resistivity is introduced in the induction equation, thereby avoiding reliance on artificially specified resistivity coefficients to trigger reconnection. By minimizing the resistivity, we effectively maximize the Lundquist number at a given grid resolution, making the simulation closer to real coronal conditions.

In this study, our simulations show that, in a stable, strongly sheared magnetic field configuration within a bipolar active region, magnetic flux diffusion can effectively drive portions of the magnetic flux toward the PIL. 
The primary role of magnetic flux diffusion is to continuously expand the core magnetic field, enhancing the pinching effect that leads to the formation of a current sheet. When magnetic reconnection occurs within this current sheet, the eruption is initiated.
Additionally, flux diffusion contributes to magnetic cancellation near the PIL, leading to the formation of a weakly twisted magnetic flux rope above the PIL. However, this flux rope plays a limited role in the initiation of the eruption, as its spatial position remains largely unchanged throughout the eruption.

Many observational studies have reported the persistence of post-eruption filaments in active region~\citep{Tang1986, Cheng2014, Dhakal2020}, supporting the interpretation that the pre-existing flux rope by magnetic cancellation is not the trigger of the eruption~\citep{Yardley2018}.
This apparent stability of this flux rope can be attributed to two factors: first, the bottom diffusion and cancellation process alone is insufficient to generate a strongly twisted flux rope capable of triggering MHD instability; second, the flux rope's position beneath the flare current sheet prevents it from rising during flare reconnection. Instead, the reconnection outflows exert a downward compression, pushing the flux rope closer to the bottom boundary and enhancing the tangential component of the bottom magnetic field~\citep{Bian2023}.

Furthermore, our results also demonstrate that the flux diffusion enhances the non-potentiality of the magnetic field. Once this non-potentiality approaches its upper limit, the magnetic system becomes unstable, and the accumulated free magnetic energy is released through an eruption.


In summary, our findings indicate that magnetic reconnection within the current sheet of core field serves as both the trigger and driver of the eruption. This mechanism is consistent with the eruption initiation mechanism identified in our previous studies~\citep{Jiang2021b, Bian2022a, Bian2022, Jiang2024}. These findings offer new insights into the physical mechanisms underlying solar eruptions, emphasizing the significance of current sheet formation in driving large scale eruption events. In future work, we plan to stop shearing earlier and start converging (flux cancellation) sooner to form a large flux MFR. The subsequent rising process of the MFR and the formation position of the current sheet will then be examined to determine whether the eruption is initiated by the ideal instability of the MFR \citep{Aulanier2010} or remains triggered and driven by current sheet reconnection \citep{Liu2024d}.

%

\begin{acks}
We grateful acknowledge NASA SDO/AIA science teams for providing observation data. 
\end{acks}

\begin{authorcontribution}
X.B. carried out the simulation and wrote the draft of the manuscript. C.J. leads this work and all contribute to the study.
\end{authorcontribution}

\begin{fundinginformation}
This work is jointly supported by National Natural Science Foundation of China (NSFC 42174200), Shenzhen Science and Technology Program (grant no. RCJC20210609104422048), Shenzhen Key Laboratory Launching Project (no. ZDSYS20210702140800001), Guangdong Basic and Applied Basic Research Foundation (2023B1515040021). The computational work was carried out on TianHe-1(A), National Supercomputer Center in Tianjin, China.
\end{fundinginformation}
\begin{dataavailability}
No datasets were generated or analyzed during the current study.
\end{dataavailability}
%
%

%
\bibliographystyle{spr-mp-sola}
\bibliography{mylib}  



\end{document}